\documentclass[%
nofootinbib,
 amsmath,amssymb,
 aps, physrev,
prd,
twocolumn
]{revtex4-1}

\usepackage[utf8]{inputenc}
\usepackage{graphicx}% Include figure files
\usepackage{dcolumn}% Align table columns on decimal point
\usepackage{bm}% bold math
\usepackage[colorlinks,linkcolor=magenta,citecolor=magenta,urlcolor=magenta]{hyperref}

\usepackage{bm,graphicx,dcolumn,epstopdf,epsf, latexsym,mathbbol, amssymb,amsmath,color,slashed, mathrsfs,mathcomp,simplewick}
\usepackage[center]{subfigure}
\usepackage{multirow}
\usepackage{graphicx}
\usepackage{makecell}
\usepackage{epstopdf}
\usepackage[table]{xcolor}
\usepackage{multirow}
\usepackage[table]{xcolor}
\usepackage{booktabs}
\usepackage{geometry}
\newcommand{\orcid}[1]{\href{https://orcid.org/#1}{\includegraphics[width=10pt]{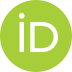}}}
\usepackage{booktabs} 
\graphicspath{{figs/}} % Specify the path to the figures folder

\begin{document}

\preprint{APS/123-QED}

\title{\textbf{Reconciling Fractional Power Potential and  EGB Gravity in the light of ACT } 
}% 
\author{Mehnaz Zahoor\orcid{0009-0009-7926-4861}}
\email{mehnazzahoormir@gmail.com}
\affiliation{Department of Physics, Central University of Kashmir, Ganderbal J\&K- 191131 }

\author{Suhail Khan\orcid{0009-0007-4941-0069}}
\email{suhail@ctp-jamia.res.in}
\affiliation{Centre for Theoretical Physics, Jamia Millia Islamia, New Delhi 110025, India}

\author{Imtiyaz Ahmad Bhat \orcid{0000-0002-2695-9709}}
\email{bhat.imtiyaz@iust.ac.in}
\affiliation{Department of Physics, Islamic University of Science and Technology, Awantipora, J\&K- 192122 }

%\affiliation{}
%  First affiliation for this author
% Anzhong_Wang@baylor.edu
 %\affiliation{
%  second institution for this author
% \affiliation{%
%  Authors' institution and/or address\\
%  This line break forced with \textbackslash\textbackslash
% }%

% \collaboration{CLEO Collaboration}%\noaffiliation


\begin{abstract}
Recent results from the ACT collaboration indicate a higher value for the scalar spectral index, with $n_s = 0.9743 \pm 0.0034$,  which sets the tighter constraints on inflationary models and these shifts are not in favor of many pre-existing scenarios, including the widely studied and accepted standard Starobinsky model. In this paper we examine the fractional power scalar potential within the framework of Einstein–Gauss–Bonnet (EGB) gravity incorporating  standard slow-roll approximation. The EGB theory, motivated by higher-dimensional models, introduces quadratic curvature corrections and a coupling between the scalar field and the Gauss–Bonnet term, thereby modifying the cosmological dynamics. \iffalse In this study, we analyze the model using both the slow-roll approximation for various fractional value of $n$ \fi The results show good agreement with observational data, placing the predictions within the $1\sigma$ region of the ACT $r$–$n_s$ constraint plot. Furthermore, incorporating the running of the scalar spectral index reinforces the model’s consistency with observational bounds. We also explore the parameter space of the EGB couplings and identify the range of free parameters for which the results of  $n_s$ and $r$ values remain within the $1\sigma$ region of the ACT constraints. Finally, we also investigate the reheating phase, demonstrating that the model not only agrees with ACT data but also satisfies the lower bound on the reheating temperature, thereby ensuring a consistent and viable cosmological scenario.

%It is determined that increasing the value of the coupling free parameters $\xi_2$ increases the scalar spectral index and the tensor-to-scalar ratio. For the $0.0012 < \xi_2 < 0.01$ range, the resulting values for $r-n_s$ stand in the $1\sigma$ region of P-ACT-BL-BK18. Then, it is concluded that the Fractional potential in the EGB gravity can satisfy the constraint of ACT data.  
\end{abstract}

\maketitle

%\tableofcontents

%%%%%%%%%%%%%%%%%%%%%%%%%%%%%%%%%%%%%%%%%%%%%%%%%%%%%%%%%%%%%
%%%%%%%%%%%%%%%%%%%%%%%%%%%%%%%%%%%%%%%%%%%%%%%%%%%%%%%%%%%%%
%%%%%%%%%%%%%%%%%%%%%%%%%%%%%%%%%%%%%%%%%%%%%%%%%%%%%%%%%%%%%
%%%%%%%%%%%%%%%%%%%%%%%%%%%%%%%%%%%%%%%%%%%%%%%%%%%%%%%%%%%%%
\section{\label{introduction}
        Introduction}
Although the inflation in the early Universe resolves many problems like horizon, flatness, and monopole problems, it anticipates quantum fluctuations that serves as the seed for the universe's large-scale  structure formation~\cite{Guth:1980zm,Linde:1981mu,Mukhanov:1981xt,Sato:1981qmu,1996tyli.conf..771S, PhysRevLett.48.1220,Starobinsky:1980te,starobinsky:1982ee,starobinsky:1983zz}. The data from observations~\cite{WMAP:2012nax, Planck:2015sxf,Planck:2018jri} collected over the past few decades strongly substantiated the model of inflation and is now the fundamental component of all cosmological models. Since the first proposal of inflationary cosmology, the model has been developed in different ways ~\cite{Barenboim:2007ii,Franche:2010yj,Unnikrishnan:2012zu,Saaidi:2015kaa,Mohammadi:2015jka,
Fairbairn:2002yp,Aghamohammadi:2014aca,
Bessada:2009pe,Weller:2011ey,Nazavari:2016yaa,Amani:2018ueu,Mohammadi:2018zkf,
berera1995warm,berera2000warm,BasteroGil:2004tg,Sayar:2017pam,Akhtari:2017mxc,Sheikhahmadi:2019gzs,Motohashi:2014ppa,Odintsov:2017hbk,Oikonomou:2017bjx,Mohammadi:2019qeu,Mohammadi:2020ftb,Mohammadi:2022vru,Adhikari:2020xcg}.  The model has been thoroughly investigated in different scenarios and in numerous modified gravity theories \cite{Mohammadi:2023kzd}. The inflationary phase is driven by a scalar field slowly rolling down a flat potential towards it minimum, which in return fuels a quasi-de Sitter expansion, leading to a rapid expansion of the universe's size over a short epoch of time.

Recent data from the Atacama Cosmology Telescope (ACT) shows that the scalar spectral index $n_s$ is on higher side as  compared to what \textit{Planck}  observations reported earlier. After a combined look at \textit{Planck} and ACT data, the scalar spectral index is measured as $n_s = 0.9709 \pm 0.0038$. By incorporating further information from CMB, BAO, and DESI, the value rises to $n_s = 0.9743 \pm 0.0034$ \cite{ACT:2025fju, ACT:2025tim}. 

This new observational discovery suggests that several widely accepted inflationary models, including the Starobinsky model~\cite{Starobinsky:1980te}, which was compatible with the earlier data, are now disfavored since they stand on the $2 \sigma$ border. This shift in the $r-{n_s}$ parameter space highlights how crucial it is to update and improve the consistent inflationary models. 
Consequently, in light of the most recent observational evidence, the wide class of inflationary models need to be revisited\cite{Kallosh:2025rni,Aoki:2025wld,Dioguardi:2025vci,Salvio:2025izr,Brahma:2025dio,Gao:2025onc,Drees:2025ngb,Zharov:2025evb,Yin:2025rrs,Liu:2025qca,Gialamas:2025ofz,Haque:2025uri,Haque:2025uis,Dioguardi:2025mpp,Gialamas:2025kef,Yogesh:2025wak,Mohammadi:2025gbu,risdianto2025,
ferreira2025,
mohammadi2025,
pallis2025b,
gao2025,
okada2025,
mcdonald2025,
bianchi2025,
odintsov2025,
kohri2025,
chakraborty2025,
pallis2025,
frolovsky2025,
haque2025,
peng2025,
bernardo2025,
yi2025,
addazi2025,
maity2025,
byrnes2025,
zharov2025}.

A reheating phase is necessary to re-populate the Universe, which is in a cold condition at the end of inflation\cite{Kofman:1994rk}. Standard particles are produced during the reheating phase by the decay of the scalar field. Interactions between the produced particles cause the universe to warm, facilitating a seamless transition to the radiation phase (for a more thorough description of the reheating, see \cite{Lozanov:2019jxc, Kofman:1997pt}). Any observation does not strictly constrain the reheating temperature.

However the BBN provides a lower range for the reheating temperature, $T_{\rm BBN} \simeq 10^{-2} \; {\rm GeV}$, while as the scale at which inflation occurs provides an upper limit, $T_{re} < 10^{16} \; {\rm GeV}$. The reheating temperature thus has a broad valid available range from inflation to BBN.

 One can get low values of the tensor-to-scalar ratio, $r$ ~~\cite{Bezrukov:2007ep,Bezrukov:2010jz}, in cosmological models, having a non-minimal coupling. But the recent observations allow a relatively small upper bound on the tensor-to-scalar ratio, $r < 0.038$ (P-ACT-LB-BK18), which has drawn attention to these models.  However, superstring theory-inspired gravitational theories with higher-order curvature terms conclude that, in addition to the higher-order curvature terms, there should be a non-minimal interaction between the scalar field and the geometry. Thus, there is a compelling need to take into account modified gravity theories in order to comprehend the early stages of evolution of our  Universe.
Even the inflation potentials that are not allowed in Einstein gravity can be reconciled by modified gravity theories~~\cite{Barvinsky:1994hx,Cervantes-Cota:1995ehs,Barvinsky:2008ia,DeSimone:2008ei,Gialamas:2020vto,Bezrukov:2008ej,Kallosh:2013pby,Barvinsky:2009ii,Rubio:2018ogq,Gialamas:2023flv,Gialamas:2022xtt,Gialamas:2021enw,Gialamas:2020snr,Kim:2025ikw,Kim:2025dyi,Yogesh:2024mpa,Yogesh:2024vcl,Koshelev:2020xby}. 

Here, we are interested in investigating the fractional power Potential in the framework of EGB gravity, which is a modified gravity theory that incorporates quadratic curvature corrections and is determined by the higher-dimensional theory. A scalar field with a non-minimal coupling to the GB component is present in the theory and is described by $\xi(\phi)$~\cite{vandeBruck:2015gjd,Guo:2009uk,Koh:2016abf,Satoh:2008ck,Jiang:2013gza,Koh:2018qcy,Mathew:2016anx,Pozdeeva:2020apf,Pozdeeva:2016cja,Nozari:2017rta,Odintsov:2018zhw,Fomin:2019yls,Odintsov:2020sqy,Odintsov:2020zkl,Kawai:2021bye,Kawai:2017kqt,Oikonomou:2022xoq,Oikonomou:2022ksx,Cognola:2006sp,Odintsov:2020xji,Nojiri:2019dwl,Fomin:2019yls,Oikonomou:2024etl,Oikonomou:2024jqv,Odintsov:2023weg,Kawai:1998ab,Nojiri:2024hau,Elizalde:2023rds,Nojiri:2023mvi,Odintsov:2023aaw,Odintsov:2022rok,Odintsov:2022rok,Kawai:2023nqs,Yogesh:2025hll,Khan:2022odn,Gangopadhyay:2022vgh}.The dynamics of the universe is mainly influenced by the Gauss-Bonnet term. In string theory, it functions as a quantum correction to the Einstein-Hilbert action. Here, we look into inflation in the EGB gravity frame using two well-motivated GB coupling functions one as hyperbolic~\cite {Kawai:2021edk,Ashrafzadeh:2023ndt,Yogesh:2024zwi,Yogesh:2024zwi} and other as exponential~\cite{Pozdeeva:2020shl,Yi:2018gse,Kleidis:2019ywv,Rashidi:2020wwg} and the form of these couplings is given implicitly in Eq. \ref{coupling_function}. It is believed that the fractional potential describes the potential of the scalar field. We find that the model shows high agreement with the ACT data, and the anticipated values of the scalar spectral index and the tensor-to-scalar ratio remain within the $1\sigma$ area, despite the potential being disfavored in the Eisenstein gravity. We also take into account the reheating phase, as well as how the reheating temperature and e-fold behave in relation to the scalar spectral index. We find that for $\omega_{re} > 1/3$, the model can both meet the reheating temperature requirement and achieve high agreement in $1\sigma$ range for different sets of model parameters in concordance with the ACT data.

Our analysis is structured as follows: we provide a quick overview of the EGB gravity theory in Sec.\ref{sec:model}. Next, we apply the slow-roll approximations to redefine the key dynamical equations and explore the inflationary and reheating phases in Sec.\ref{sec:inflation}. The ACT data is compared with the model's output from the slow-roll approach in Sec.\ref{sec:result} for hyperbolic coupling and in Sec.\ref{expc} for exponential coupling respectively. Lastly, we provide a summary of our analysis in Sec.\ref{sec:conclusion}. \\

%%%%%%%%%%%%%%%%%%%%%%%%%%%%%%%%%%%%%%%%%%%%%%%%%%%%%%%%%%%%%
%%%%%%%%%%%%%%%%%%%%%%%%%%%%%%%%%%%%%%%%%%%%%%%%%%%%%%%%%%%%%
%%%%%%%%%%%%%%%%%%%%%%%%%%%%%%%%%%%%%%%%%%%%%%%%%%%%%%%%%%%%%
%%%%%%%%%%%%%%%%%%%%%%%%%%%%%%%%%%%%%%%%%%%%%%%%%%%%%%%%%%%%%
%%%%%%%%%%%%%%%%%%%%%%%%%%%%%%%%%%%%%%%%%%%%%%%%%%%%%%%%%%%%%
%%%%%%%%%%%%%%%%%%%%%%%%%%%%%%%%%%%%%%%%%%%%%%%%%%%%%%%%%%%%%
\section{\label{sec:model}
 EGB Background}
The action of the model is given by~\cite{Pozdeeva:2020apf}
\begin{equation}
\label{action1}
S=\int d^4x\sqrt{-g}\left[ \frac{R}{2} - \frac{1}{2}g^{\mu\nu}\partial_\mu\phi\partial_\nu\phi-V(\phi)-\frac{1}{2}\xi(\phi){\cal G}\right],
\end{equation}
where
 \begin{equation}
     \mathcal{G}=R_{\mu\nu\rho\sigma}R^{\mu\nu\rho\sigma}-4R_{\mu\nu}R^{\mu\nu}+R^2, \nonumber
 \end{equation}
is the quadratic curvature correction term also known as GB term, $g$ is the determinant of the metric $g_{\mu\nu}$, $R,R_{\mu\nu},R_{\mu\nu\rho\sigma}$ is the Ricci scalar, Ricci tensor, Riemann tensor of corresponding metric $g_{\mu\nu}$ respectively, and $V(\phi)$ is the potential of the scalar field $\phi$. 

Assuming homogeneity and isotropy, the large-scale geometry of the universe is described by a spatially flat FLRW metric. Under this assumption, the modified Friedmann equations can be derived accordingly.
% Assuming that the geometry of the universe is described by a spatially flat FLRW metric, 
% \begin{equation}
%     ds^2 = -dt^2 + a^2(t) \big( dx^2 + dy^2 + dz^2 \big),
% \end{equation}
% where $a(t)$ is the scale factor of the universe, the modified Friedmann equations are given as
\begin{eqnarray}
6H^2\left(1-4\xi_{,\phi} \dot\phi H\right)&=&\dot\phi^2+2V,
\label{Equ00} \\
2\dot H\left(1 - 4\xi_{,\phi}\dot\phi H\right)&=&{}-\dot\phi^2+4H^2 \; \Psi,
\label{Equ11}
\end{eqnarray}
and for the dynamics scalar field, the equation is written as
\begin{equation}
\ddot\phi+3H\dot\phi = {} -V_{,\phi} -12H^2\xi_{,\phi}\left(\dot{H}+H^2\right),\label{Equphi}
\end{equation}
Here, $\Psi$ is defined as
\[
\Psi = \xi_{,\phi\phi}\dot\phi^2 + \xi_{,\phi}\ddot\phi - H\xi_{,\phi}\dot\phi,
\]
where $H = \dot{a}/a$ is the Hubble parameter, and \emph{dot} denotes a derivative with respect to cosmic time. In case of a varying coupling function, i.e., $\xi_{,\phi} \neq 0$, Eqs.~\eqref{Equ11} and \eqref{Equphi} do not represent the full dynamical system. The correct dynamical equation can be derived from Eqs.~\eqref{Equ00} and \eqref{Equphi} and performing algebraic transformations; for further details, see Ref.~\cite{Pozdeeva:2024ihc}.

\begin{widetext}
\begin{equation}
\label{DynSYSN}
\begin{split}
\frac{d\phi}{dN}=&\,\chi,\\
\frac{d\chi}{dN}=&\,\frac{1}{H^2\left(B-2\xi_{,\phi}H^2\chi\right)} \Big(
3\left[3-4\xi_{,\phi\phi} H^2\right]\xi_{,\phi}H^4\chi^2+ \left[3B+2\xi_{,\phi}V_{,\phi}-3\right]H^2\chi-4V^2 X \Big) -\frac{\chi}{2H^2}\frac{dH^2}{dN},\\
\frac{dH^2}{dN}=&\,\frac{H^2}{2\left(B-2\xi_{,\phi}H^2\chi\right)} \Big( \left(4\xi_{,\phi\phi}H^2-1\right)\chi^2-16\xi_{,\phi}H^2\chi-16 V^2 \xi_{,\phi}X \Big).
\end{split}
\end{equation}  
\end{widetext}

Here, $\chi$ is defined as $\chi = \dot{\phi} / H$, and a change of variables has been applied to express the equations in terms of the number of e-folds, $N = \ln(a / a_e)$, rather than cosmic time $t$, using the relation $dN = H\,dt$. The quantities $B$ and $X$ are defined as follows:

\begin{equation}
B=12\xi_{,\phi}^2H^4 + \frac{1}{2},\qquad
X=\frac{1}{4V^2}\left(12\xi_{,\phi} H^4+V_{,\phi} \right).
\end{equation}

% \begin{equation}
% \label{DynSYS}
% \begin{split}
% \dot\phi=&\psi,\\~~~~~~~~~~~~~
% \dot\psi=&\frac{1}{B-2\xi_{,\phi}H\psi}\left\{
% 3\left[3-4\,\xi_{,\phi\phi} H^2 \right]\xi_{,\phi}H^2\psi^2+\left[3B+2\,\xi_{,\phi}V_{,\phi}-6U_0\right]H\psi-\frac{V^2}{U_0}X\right\},\\
% \dot H=&\frac{1}{4\left(B-2\,\xi_{,\phi}H\psi\right)}\left\{\left(4\,\xi_{,\phi\phi}H^2-1\right)\psi^2-16\xi_{,\phi}H^3\psi-4\frac{V^2}{U_0^2}\,\xi_{,\phi} H^2 X\right\},
% \end{split}
% \end{equation}\\
% where the used quantities $B$ and $X$ are defined as
% \begin{equation}
% B=12\xi_{,\phi}^2H^4+U_0,\qquad
% X=\frac{U_0^2}{V^2}\left(12\xi_{,\phi} H^4+V_{,\phi} \right).
% \end{equation}
% To consider the inflationary phase, it is a better choice to work with the number of e-fold, $N = \ln\big( a / a_e \big)$ than the cosmic time. Doing a variable change, $dN = H dt$, the dynamical equations are rewritten as 
% \begin{equation}
% \label{DynSYSN}
% \begin{split}
% \frac{d\phi}{dN}=&\,\chi,\\
% \frac{d\chi}{dN}=&\,\frac{1}{H^2\left(B-2\xi_{,\phi}H^2\chi\right)}\left\{
% 3\left[3-4\xi_{,\phi\phi} H^2\right]\xi_{,\phi}H^4\chi^2+ \left[3B+2\xi_{,\phi}V_{,\phi}-6U_0\right]H^2\chi-\frac{V^2}{U_0}X\right\}\\
% &{}-\frac{\chi}{2H^2}\frac{dH^2}{dN},\\
% \frac{dH^2}{dN}=&\,\frac{H^2}{2\left(B-2\xi_{,\phi}H^2\chi\right)}\left\{\left(4\xi_{,\phi\phi}H^2-1\right)\chi^2-16\xi_{,\phi}H^2\chi-4\frac{V^2}{U_0^2}\xi_{,\phi}X\right\}.
% \end{split}
% \end{equation}
% in which we have defined $Q\equiv H^2$ and $\chi=\psi/H$. 
The dynamics of the inflationary phase are characterized by the slow-roll parameters, which are defined as:

\begin{equation}
\label{epsilon}
\varepsilon_1 ={}-\frac{\dot{H}}{H^2}={}-\frac{d\ln(H)}{dN},\qquad \varepsilon_{i+1}= \frac{d\ln|\varepsilon_i|}{dN},\quad i\geqslant 1,
\end{equation}

In EGB gravity, there are additional slow-roll parameters in addition to the above-mentioned parameters in Eq.\eqref{epsilon}. These parameters are however, associated with the coupling term, and are defined as:
\begin{equation}
\label{delta}
\delta_1 = %\frac{2}{U_0}\xi_{,\phi}H\dot\phi=
4\xi_{,\phi}H^2\chi,\qquad \delta_{i+1}=\frac{d\ln|\delta_i|}{dN}, \quad i\geqslant 1.
\end{equation}
In terms of these slow-roll parameters, the evolution equations get redefined as discusses in   Ref.~\cite{Pozdeeva:2024ihc}. Consequently, both the scalar spectral index and the tensor-to-scalar ratio can also be expressed in terms of these slow-roll parameters which are now dependent on coupling term and are given as:

\begin{equation}
\label{ns_slr}
n_s=1-2\varepsilon_1-\frac{2\varepsilon_1\varepsilon_2-\delta_1\delta_2}{2\varepsilon_1-\delta_1}, \qquad r=8|2\varepsilon_1-\delta_1|. 
%=1-2\varepsilon_1-\frac{d\ln(r)}{dN}=1+\frac{d}{dN}\ln\left(\frac{Q}{U_0r}\right),
\end{equation}
% \begin{equation}
% \label{r_slr}
% r=8|2\varepsilon_1-\delta_1|.
% \end{equation}

%%%%%%%%%%%%%%%%%%%%%%%%%%%%%%%%%%%%%%%%%%%%%%%%%%%%%%%%%%%%%
%%%%%%%%%%%%%%%%%%%%%%%%%%%%%%%%%%%%%%%%%%%%%%%%%%%%%%%%%%%%%
%%%%%%%%%%%%%%%%%%%%%%%%%%%%%%%%%%%%%%%%%%%%%%%%%%%%%%%%%%%%%
%%%%%%%%%%%%%%%%%%%%%%%%%%%%%%%%%%%%%%%%%%%%%%%%%%%%%%%%%%%%%
%%%%%%%%%%%%%%%%%%%%%%%%%%%%%%%%%%%%%%%%%%%%%%%%%%%%%%%%%%%%%
%%%%%%%%%%%%%%%%%%%%%%%%%%%%%%%%%%%%%%%%%%%%%%%%%%%%%%%%%%%%%
\section{\label{sec:inflation}
        Inflation and subsequent reheating phases}

This section considers the inflationary and reheating phases within the EBG background. The potential under consideration is fractional power law originally proposed by~\cite{Harigaya_2014}, though disfavored in the context of standard gravity, but can be rescued in EGB  when confronted with recent ACT data. 
\begin{equation}\label{potential}
    V(\phi) = V_0 \;\phi^{n}
\end{equation}
where the power index $n$ is set to ($1/3, 2/5, 2/3, 4/3$). The choice of $n$ values is motivated from different  literature surveys and observational data. The choice for  $n =2/3 ~ \text{and} ~ 4/3$ are considered in the context of the \textit{Planck} results, $n =1/3 ~2/5 ~ \text{and} ~ 2/3$ are inspired from Ref.\cite{Ashrafzadeh_2024}  and $n =1/3 ~ \text ~ 2/3$  are further supported from ACT DR6 constraints on extended cosmological models.

Two commonly used forms of the Gauss--Bonnet (GB) coupling function are considered \cite{Khan:2022odn,Gangopadhyay:2022vgh,Yogesh:2024mpa,Pozdeeva:2020shl,Jiang:2013gza,Yi:2018gse,Odintsov:2018zhw,Kleidis:2019ywv,Rashidi:2020wwg} and defined as follows:
 
\begin{equation}\label{coupling_function}
    \xi(\phi) = \frac{\xi_1}{V_0} \;  \tanh(\xi_2 \; \phi), \quad 
    \xi(\phi) = \frac{\xi_1}{V_0} \; e^{-\xi_2 \; \phi}
\end{equation}
 Here, $\xi_1$ and $\xi_2$ are free coupling parameters. The constant $V_0$ is introduced to simplify subsequent calculations and does not affect the generality of the analysis, because most of the slow roll parameters are independent of $V_{0}$. This work follows the effective potential method developed in~\cite{Pozdeeva:2019agu, Pozdeeva:2020apf, Vernov:2021hxo}, where the stability of de Sitter solutions in the model has been investigated. The effective potential is expressed in terms of the scalar field potential and the coupling function as follows:
\begin{equation}
\label{Veff}
V_{eff}(\phi)={}-\frac{1}{4V(\phi)}+\frac{1}{3}\xi(\phi).
\end{equation}
 It is assumed that in the standard slow-roll approximation, the parameters defined in Eqs.~\eqref{epsilon} and \eqref{delta} are  small. This kind of assumption, known as the slow-roll approximation, allows for a simplified treatment of the inflationary dynamics.
Consequently, the  inflationary dynamical equations under such assumptions  are simplified, and it leads to;
\begin{equation}
\label{N1}
\frac{dN}{d\phi}\simeq{}-\frac{1}{4V{V_{eff}}_{,\phi}}\,.
\end{equation}

By solving the integral, the scalar field is obtained as a function of the number of e-folds $N$. Additionally, the Hubble parameter and the quantity $\chi$ are expressed in terms of the scalar potential and the GB coupling term as :
\begin{equation}\label{EquSsr}
H^2 \simeq \frac{V(\phi)}{3}, \qquad 
\chi\simeq {}-4{V_{eff}}_{,\phi}.
\end{equation}
By applying the approximations, the slow-roll parameters are expressed in terms of the scalar field. Substituting the solution obtained from Eq.~\eqref{N1} into the slow-roll parameters, the scalar spectral index and the tensor-to-scalar ratio are then derived using Eq.~\eqref{ns_slr}, and hence can be expressed as the functions of the number of e-folds $N$. \\

After the end of inflation, the decay of the scalar field’s energy into standard model particles leads to the re-population of the universe and the onset of a hot, radiation-dominated era. Since reheating lacks direct observational signatures, it is useful to study this phase indirectly by relating the reheating temperature to inflationary observables~\cite{Cook:2015vqa}. Assuming that the matter content during reheating can be characterized by a constant equation of state parameter $\omega_{re}$, one can derive the following relation~\cite{Cook:2015vqa,Khan:2022odn,Gangopadhyay:2022vgh,Adhikari:2019uaw,Gialamas:2019nly,Yogesh:2024vcl,Yogesh:2024iip,Adhikari:2020xcg,Adhikari:2019uaw}:

\begin{equation}
N_{re}= \frac{4}{ (1-3w_{re} )}   \left[61.488  - \ln \left(\frac{ V_{end}^{\frac{1}{4}}}{ H_{k} } \right)  - N_{k}   \right]
\label{Nre}
\end{equation}
\begin{equation}
T_{re}= \left[ \left(\frac{43}{11 g_{re}} \right)^{\frac{1}{3}}    \frac{a_0 T_0}{k_{}} H_{k} e^{- N_{k}} \left[\frac{3^2 \cdot 5 V_{end}}{\pi^2 g_{re}} \right]^{\beta}  \right]^\gamma,
\label{Tre}
\end{equation} 
Here, $T_{re}$ and $N_{re}$ denote the reheating temperature and the number of e-folds during reheating, respectively. The quantity $N_k$ represents the number of e-folds from the horizon crossing of a given scale to the end of inflation. The constant parameters $\beta$ and $\gamma$ are defined as $\beta = -\frac{1}{3}(1 + \omega_{re})$ and $\gamma = \frac{\beta}{1 - 3\omega_{re}}$, respectively.

\section{\label{sec:result}
         Case I: Hyperbolic Coupling }%Consistency with the ACT data}
The consistency of the model with observational data is investigated using slow-roll approximation Eq. ~\eqref{epsilon},~\eqref{delta}. These analyses are performed for the two coupling functions introduced in Eq.~\eqref{coupling_function}.
%%%%%%%%%%%%%%%%%%%%%%%%%%%%%%%%%%%%%%%%%%%%%%%%%%%%%%%%%%%%%
%%%%%%%%%%%%%%%%%%%%%%%%%%%%%%%%%%%%%%%%%%%%%%%%%%%%%%%%%%%%%
%%%%%%%%%%%%%%%%%%%%%%%%%%%%%%%%%%%%%%%%%%%%%%%%%%%%%%%%%%%%%
% \subsection{Hyperbolic coupling}

 %Repeating the same process for the new slow-roll approximations I and II, one can find the corresponding $n_s$ and $r$. 
As a first case, it is assumed that the  GB coupling is described by a $\textit{tanh}$ function, as given in Eq.~\eqref{coupling_function}. The analysis begins with the standard slow-roll approximation, which requires solving the integral in Eq.~\eqref{N1}. It is assumed that the number of e-folds at horizon crossing is $N_\star = 0$, and that inflation ends at $N = 60$. The initial condition is set at the end of inflation, where the first slow-roll parameter satisfies $\varepsilon_1(\phi_e) = 1$, allowing the determination of the field value at the end of inflation. Solving Eq.~\eqref{N1} yields the scalar field as a function of the number of e-folds during inflation. This solution is then used to compute the slow-roll parameters, the scalar spectral index $n_s$, and the tensor-to-scalar ratio $r$ as functions of the number of e-folds $N$.
\begin{figure}[htb]
    \centering
    \includegraphics[width=1.0\linewidth]{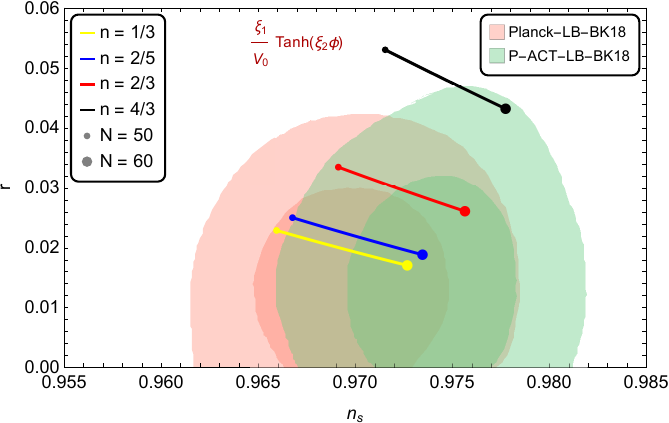}
    \caption{The plot illustrates the predictions of the model in the $r - n_s$ plane using the standard slow-roll approximation for the case of hyperbolic coupling. The predictions are compared against the ACT observational constraints in the $r – n_s$ plane for various values of the fractional value $n$, while keeping the coupling fixed at $\xi_1 = 15, \; \xi_2 = 0.255$. Legends are self explanatory.}
    \label{rvnstan}
\end{figure}

Figure~\ref{rvnstan} presents the $r - n_s$ plot obtained from the slow-roll approximation for different values of the  power $n$ of the scalar field potential. The larger (smaller) markers correspond to results for $N = 60$ ($N = 50$) e-folds. It is observed that for the values of $n =1/3, ~ 2/5~ $, and setting e-folds to  $N=60$ the prediction lies  well within the $1\sigma$ of the P-ACT-LB-BK18 constraints. While for $n=4/3$ for $60$ e-folds the results lies in the $2~\sigma$ region of P-ACT-LB-BK18, but lies outside the \textit{Planck} results, Clearly hints to revisit the cosmology models. We set coupling values to $\xi_1=15$ and $\xi_2=0.255$, and the the inflation observables $n_s$ and $r$ for all the particular choice of $\xi_1$, $\xi_2$ and $n$ for $50$ and $60$ e-folds is mentioned in table \ref{tab:coupling-clear}. All these results are well within the P-ACT-LB-BK18 dataset, indicating good agreement with ACT observations. 

From Fig.~\ref{rvnstan}, it can be seen that increasing the value of $n$ results in rising values of both $n_s$ and $r$ as mentioned in table \ref{tab:coupling-clear}. If this trend continues, the predictions eventually fall outside the $2\sigma$ bounds of both observational datasets. Similarly, increasing $n$ beyond a certain threshold also pushes the predictions for $n_s$ and $r$ beyond the observationally allowed region. These trends highlight the sensitivity of the model predictions to the choice of the fractional power $n$ of scalar field potential. However, deviations from this optimal value of $n$ or increasing cause the predictions to gradually shift outside the observational bounds of ACT.

\begin{figure}[htb]
    \centering
    \includegraphics[width=1\linewidth]{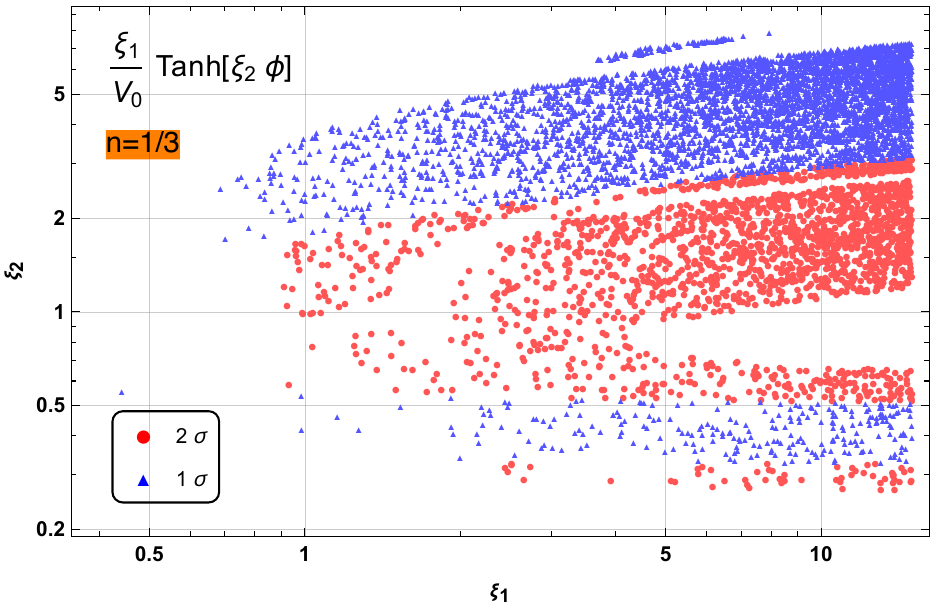}
    \caption{Hyperbolic Coupling:- The plot shows the allowed region in the $\xi_1$–$\xi_2$ parameter space for the case $n = 1/3$. The blue triangular points denote values of $(\xi_1, \xi_2)$ that yield $(n_s, r)$ predictions within the $1\sigma$ confidence region of the ACT data. The red circular points correspond to values lying in the $2\sigma$ region but outside the $1\sigma$ bounds.}
    \label{tanxi}
\end{figure}

Fig.~\ref{tanxi} illustrates the parameter space of the free parameters $\xi_1$ and $\xi_2$. For the analysis we scanned the parametric space from  0.001 to 15 for both  $\xi_1$ and $\xi_2$ . The blue dots represent combinations of $(\xi_1, \xi_2)$ for which the resulting values of $n_s$ and $r$ lie within the $1\sigma$ confidence region of the ACT data. The red circular points correspond to values lying in the $2\sigma$ region but outside the $1\sigma$ bounds. This  figure demonstrates the sensitivity of the inflationary predictions to the values of the free parameters. However we can not directly or indirectly put some bound from this analysis on these free parameters of couplings in both the cases ( Tanh and Exp).

It is observed that small values of $\xi_1 \; \text{and} \; \xi_2 $  almost $ < 1$ are not favored. Within this region and sparse set of  ($\xi_1,\xi_2$) data points satisfies inflationary observables in $1 \sigma $  or $2 \sigma$, contrary for higher values of $\xi_1 \; \text{and} \; \xi_2 $ significant parametric space satisfies the observational constraints  for the said mentioned confidence level of inflationary observables. For consistency with the available inflationary ACT data, this suggests a preference for higher values of  non-minimal coupling parameters.

The running of the scalar spectral index  given as  $\alpha_s = {d n_s}/{d\ln k}$, serves as an additional observational constraint for evaluating the viability of an inflationary model. For the parameter values $\xi_1 = 15$, $\xi_2 = 0.255$, the evaluated values of $\alpha_s$ for specific  values of $n$ for hyperbolic coupling $\frac{\xi_1}{V_0} \rm Tanh(\xi_2 \phi)$ are :
\[
\left\{
\begin{aligned}
n = \tfrac{1}{3} &\Rightarrow \alpha_s = -0.000546025, \\
n = \tfrac{2}{5} &\Rightarrow \alpha_s = -0.000552645, \\
n = \tfrac{2}{3} &\Rightarrow \alpha_s = -0.000534904, \\
n = \tfrac{4}{3} &\Rightarrow \alpha_s = -0.000499958.
\end{aligned}
\right.
\]
Notably for high regime of one of the $\xi$, favors the $n_s$ and $r$ results within the observational constraints. The results for $\alpha_s$  still lie  inside the bounds imposed by latest  ACT observations.                                                          %These results fall inside the bounds imposed by ACT data.

    \begin{table*}[htb]
\centering
\scriptsize
\renewcommand{\arraystretch}{1.8}
\tabcolsep=8pt

% \rowcolors{3}{white}{white}
\begin{tabular}{|c|c|
>{\columncolor{blue!5}}c>{\columncolor{blue!5}}c|
>{\columncolor{blue!5}}c>{\columncolor{blue!5}}c|
>{\columncolor{blue!5}}c>{\columncolor{blue!5}}c|
>{\columncolor{blue!5}}c>{\columncolor{blue!5}}c|}
\hline
% \rowcolor{gray!10}
\multirow{2.5}{*}{\textbf{Coupling Type}} & \multirow{2}{*}{\textbf{$N$}} 
& \multicolumn{2}{>{\columncolor{blue!10}}c|}{\textbf{$n = \tfrac{1}{3}$}} 
& \multicolumn{2}{>{\columncolor{blue!10}}c|}{\textbf{$n = \tfrac{2}{5}$}} 
& \multicolumn{2}{>{\columncolor{blue!10}}c|}{\textbf{$n = \tfrac{2}{3}$}} 
& \multicolumn{2}{>{\columncolor{blue!10}}c|}{\textbf{$n = \tfrac{4}{3}$}} \\
% \rowcolor{gray!0}
& & \textbf{$n_s$} & \textbf{$r$} & \textbf{$n_s$} & \textbf{$r$} & \textbf{$n_s$} & \textbf{$r$} & \textbf{$n_s$} & \textbf{$r$} \\
\hline

% Tanh coupling section
\multirow{2}{*}{$\dfrac{\xi_1}{V_0} \tanh(\xi_2 \phi)$} 
& 50 
& 0.96592 & 0.02290 
& 0.96674 & 0.02503 
& 0.96910 & 0.03353 
& 0.97155 & 0.05300 \\
& 60 
& 0.97265 & 0.01707 
& 0.97344 & 0.01885 
& 0.97562 & 0.02610 
& 0.97772 & 0.04319 \\
\hline

% Exp coupling section
\multirow{2}{*}{$\dfrac{\xi_1}{V_0} \exp(-\xi_2 \phi)$} 
& 50 
& 0.97329 & 0.02011 
& 0.97228 & 0.02368 
& 0.96693 & 0.03373 
& 0.92793 & 0.01333 \\
& 60
& 0.97738 & 0.01618 
& 0.97648 & 0.01897 
& 0.97155 & 0.02615 
& 0.93218 & 0.00699 \\
\hline
\end{tabular}
\caption{\scriptsize Comparison of $n_s$ and $r$ for different values of fractional potential power $n$ and at different e-fold $N$ under two different  coupling types: Tanh hyperbolic and Exponential.}
\label{tab:coupling-clear}
\end{table*}

\begin{figure}[h!]
    \centering
    \includegraphics[width=9cm]{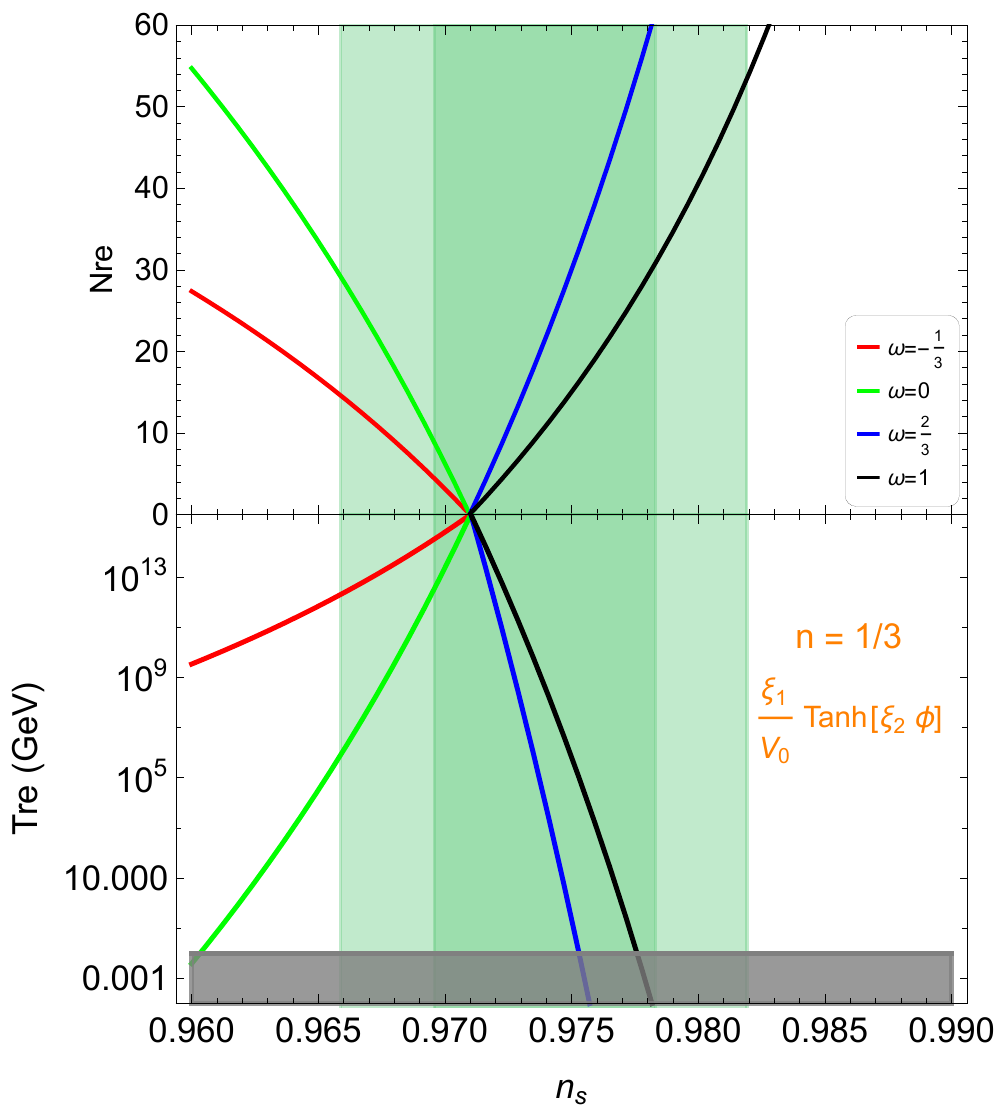}
    \caption{The plot shows the relation of $n_s$ with the reheating phase, so that the top panel shows the behavior of the reheating e-fold $N_{re}$ versus $n_s$, and the bottom panel describes the behavior of the reheating temperature $T_{re}$ versus $n_s$. The result provided for different choice of the reheating equation of state $\omega_{re}$, as $\omega_{re} = -1/3$ (red-curve), $\omega_{re} = 0$ (green-curve), $\omega_{re} = 2/3$ (blue-curve), $\omega_{re} = 1$ (black-curve). The gray area at the bottom is the excluded region originating from the BBN constraint.}
    \label{tanh_tre}
\end{figure}

After the end of inflation, the Universe undergoes a reheating phase that ensures a smooth transition to the radiation era. In Fig.~\ref{tanh_tre} we have shown the behavior of $N_{re}$ and $T_{re}$ with the spectral index $n_s$. To obtain this plot we have used  Eqs.~\eqref{Nre} and \eqref{Tre}, where the free parameters of hyperbolic coupling are chosen to be $\xi_1 = 15$ and $\xi_2 = 0.255$.

The top panel illustrates the influence of $N_{\text{re}}$ on the scalar spectral index ($n_s$). As shown in Fig.~\ref{rvnstan}, for the above mentioned values of $\xi_1$ and $\xi_2$, parameters, the scalar spectral index is found to be $n_s \simeq 0.9726$ for $N_k = 60$ and for power of potential  $n=1/3$. This value of $n_s$ corresponds to reheating e-fold numbers $N_{\text{re}} \simeq 11.57$ and $5.78$ for $\omega_{\text{re}} = 2/3$ and $1$, respectively. The bottom panel is the variation of the reheating temperature as a function of the scalar spectral index.
It is observed that higher values of the reheating equation-of-state parameter $\omega_{\text{re}}$ correspond to smaller reheating e-folds $N_{\text{re}}$, which in turn result in higher reheating temperatures. For example, for $\omega_{\text{re}} = 2/3$ and $1$, the reheating temperature is found to be $T_{\text{re}} \simeq 2.33 \times10^{9} \; \mathrm{GeV}$ and $6.62 \times10^{11} \; \mathrm{GeV}$, respectively. It is crucial to emphasize that the obtained reheating temperature must comply with both the constraints from ACT data and the lower bound set by Big Bang Nucleosynthesis (BBN), $T_{\text{re}} \gtrsim 10 \; \mathrm{MeV}$.
%%%%%%%%%%%%%%%%%%%%%%%%%%%%%%%%%%%%%%%%%%%%%%%%%%%%%%%%%%%%%
\section{Case II: Exponential coupling}\label{expc}
\begin{figure}[htb]
    \centering
    \includegraphics[width=1.0\linewidth]{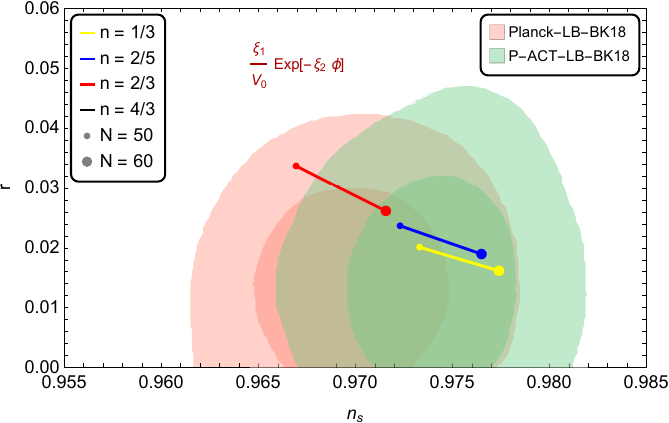}
    \caption{The plot illustrates the predictions of the model in the $r$–$n_s$ plane using the standard slow-roll approximation for the case of \textit{exponential} coupling. The predictions are compared against the ACT observational constraints in the $r$–$n_s$ plot for various values of the fractional  value $n$, while keeping the coupling fixed at $\xi_1 = 0.5,\xi_2 = 0.01$. Legends are self explanatory.}
    \label{rvsns-exp}
\end{figure}

\begin{figure}[htb]
    \centering
    \includegraphics[width=1.0\linewidth]{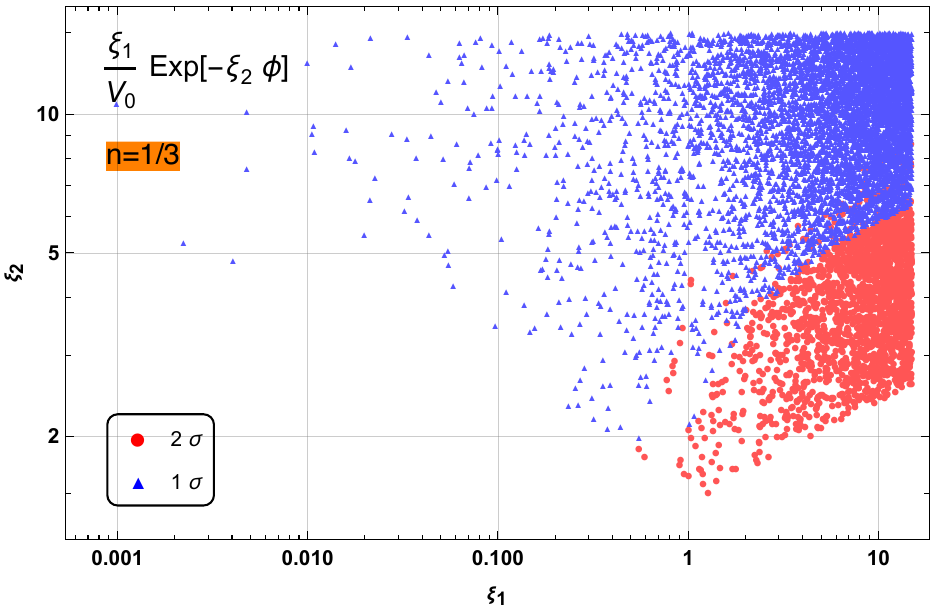}
    \caption{Exponential Coupling:- The plot shows the allowed region in the $\xi_1$–$\xi_2$ parameter space for the case $n = 1/3$. The blue triangular points denote values of $(\xi_1, \xi_2)$ that yield $(n_s, r)$ predictions within the $1\sigma$ confidence region of the ACT data. The red circular points correspond to values lying in the $2\sigma$ region but outside the $1\sigma$ bounds.}
    \label{expxi}
\end{figure}

 In case II, the consistency of this model is tested with the observational data investigated using the standard slow-roll approximations taking in consideration the second GB coupling as mention in Eq.\eqref{coupling_function}. In fig.~\ref{rvsns-exp} the yellow, blue, red lines represents the $n_s$ and $r$ dynamics while going from  50 to 60 e-folds, with $n$ $1/3,~2/5~ \text{and} ~2/3$ values of the fractional power $n$ respectively and the values for $n=4/3$ lies outside of this $r-n_s$ plane.\footnote{For $n=4/3$ the data points lies outside the $n_s-r$ plane. For $N=50$ $n_s~=~0.92793$ and $r= 0.01333$ and for  $N=60$ $n_s~=0.93218 $ and  $r=  0.00699$.} It is observed that increasing $n$ leads to a shift towards the lower side of scalar spectral index $n_s$ and substantial increase in the tensor-to-scalar ratio $r$.  For $\xi_1=0.5$  and $\xi_2=0.01$, for $n=1/3 $ and $2/5$ the values of $n_s$ and $r$ lie in the $1 \sigma $ contour of P-ACT-LB-BK18 dataset. However, as the value of $n$ increases  eventually the $n_r$ and $r$ points  move out  $1\sigma$ region. Even for the mentioned free coupling parameters, for $n=2/3$ and $N=50$ the inflationary observables lie outside the of P-ACT-LB-BK18 contour. The $r-n_s$ trajectories derived from the slow-roll approximation in Fig.\eqref{expxi} are subjected to change on changing free coupling parameters $\xi_1$ and $\xi_2$. 

%%%%%%%%%%%%%%%%%%%%%%%%%%%

\begin{figure}[htb]
    \centering
    \includegraphics[width=9cm]{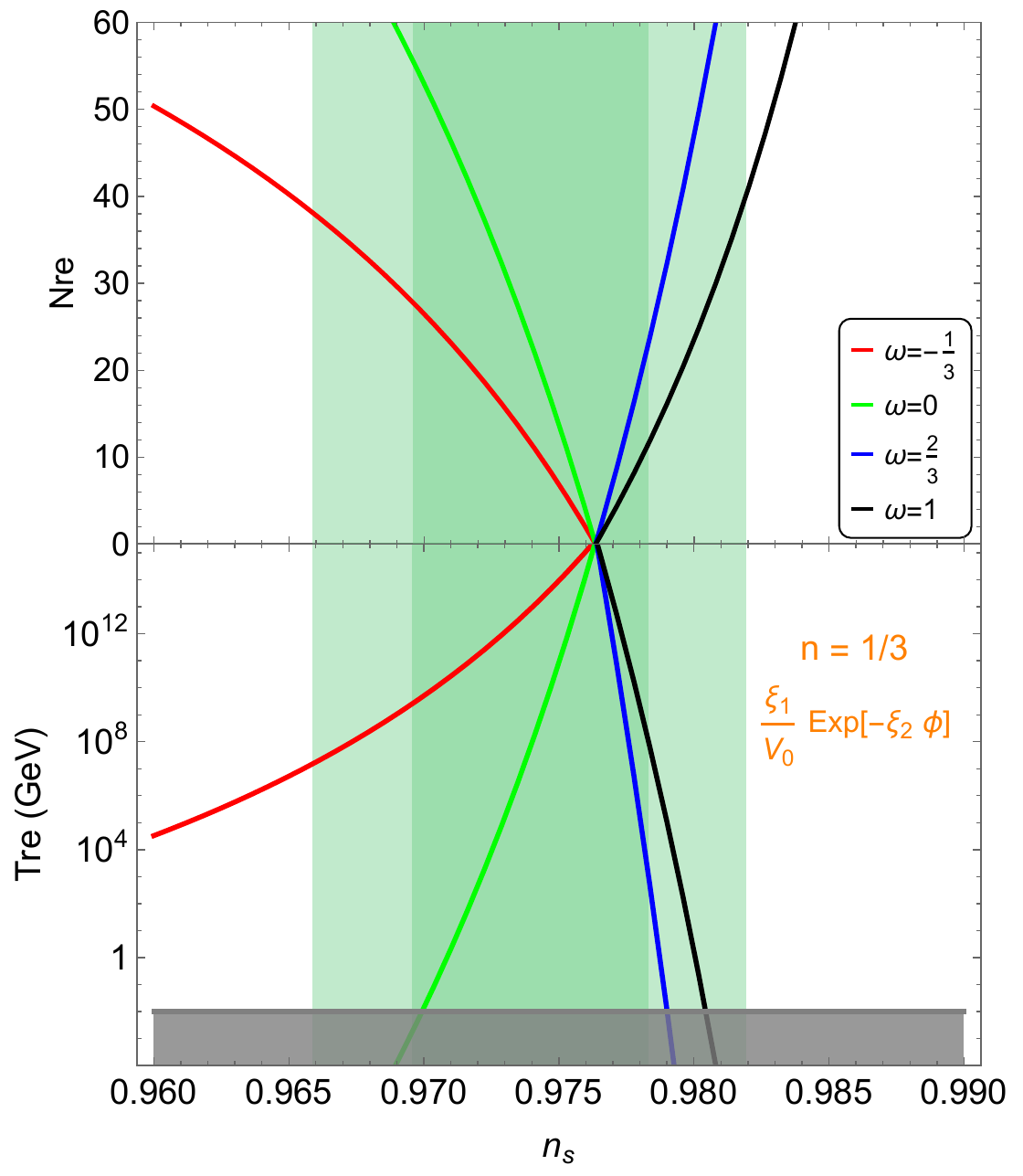}
    \caption{The relation of $n_s$ with the reheating number of e-folds $N_{re}$ is shown in the top panel, and the bottom panel displays the relation between $n_s$ and the reheating temperature $T_{re}$. The plot includes the results for different values of the reheating equation of state parameter as: $\omega_{re} = -1/3$ (red-curve), $\omega_{re} = 0$ (green-curve), $\omega_{re} = 2/3$ (blue-curve), $\omega_{re} = 1$ (black-curve). The grey area at the bottom is the excluded region originating from the BBN constraint. }
    \label{exp_tre}
\end{figure}
%%%%%%%%%%%%%%%%%%%%%%%%%%%
 The analysis for Fig.\ref{expxi} is carried out in a same way as for Fig.\ref{tanxi} for the same values of $\xi_1$ and $\xi_2$ from 0.001 to 15. In this plot, blue dots represent the combinations of $(\xi_1, \xi_2)$ for which the predicted values of $n_s$ and $r$ lie within the $1\sigma$ confidence level. The red points corresponds to the parameter sets ($n_s$ ,$r$) those which are consistent with the broader $2\sigma$ region but outside the $1 \sigma$ region. The allowed interval for $\xi_2$ depends sensitively on the choice of $\xi_1$. For instance, when $\xi_1 = 10$, the model remains compatible with the $1\sigma$ region of ACT observations provided $5 < \xi_2 \leq 15$. This permissible range of $\xi_2$ expands with increase in $\xi_1$ value. However the narrow $\xi$-regime is least favored to satisfy inflationary observables. The density of points in both $1 \sigma$ and $2 \sigma$  is more and favors higher values of free coupling parameters. But for $1 \sigma$ the permissible range of $\xi_2$ narrows on both side around the $\xi_1 = 1$. For example, at $\xi_1 = 1$, the valid range expands to $2 < \xi_2 < 15$, bot for $\xi_1 = 10$, the valid range expands to $5.5 < \xi_2 \leq 15$. Similar trend is followed for $2 \sigma $ region.

%%%%%%%%%%%%%%%%%%%%%%%%%%%
Taking into account the running of the scalar spectral index, it is found that for $\xi_1 = 0.5$, $\xi_2 = 0.01$, the evaluated values of $\alpha_s$ for the exponential coupling $\frac{\xi_1}{V_0} \exp(-\xi_2 \phi)$ are:
\[
\left\{
\begin{aligned}
n = \tfrac{1}{3} &\Rightarrow \alpha_s = -0.000341584, \\
n = \tfrac{2}{5} &\Rightarrow \alpha_s = -0.000350776, \\
n = \tfrac{2}{3} &\Rightarrow \alpha_s = -0.000386625, \\
n = \tfrac{4}{3} &\Rightarrow \alpha_s = -0.000339927.
\end{aligned}
\right.
\]
The results for running of the scalar spectral index we show here are compatible with the recent ACT observations.

The results of the reheating phase analysis are presented in Fig.~\ref{exp_tre}, which illustrates the influence of the reheating e-fold number and reheating temperature on the scalar spectral index. The plot shows how both $N_{\text{re }}$ and $T_{\text{re}}$ vary with $n_s$ for different values of the rehea ting equation-of-state parameter $\omega_{\text{re}}$, with the free parameters fixed at $\xi_1 = 0.5$ and $\xi_2 = 0.01$ and for fractional power $n=1/3$. As inferred from Fig.~\ref{rvsns-exp}, for these parameter choices, the scalar spectral index for $N_k=60$ is approximately $n_s = 0.97738$ lies in $2\sigma$ region. This value corresponds to $N_{\text{re}}\simeq 11.51$ and $5.75$ for $\omega_{\text{re}} = 2/3$ and $1$, respectively. As $\omega_{\text{re}}$ increases, the reheating e-fold decreases, leading to a higher reheating temperature, as depicted in the bottom panel. The resulting reheating temperatures are roughly $T_{\text{re}} \simeq 2.39 \times 10^{9} \; \mathrm{GeV}$ and $6.62 \times10^{11} \; \mathrm{GeV}$ for the respective values of $\omega_{\text{re}}$.

%%%%%%%%%%%%%%%%%%%%%%%%%%

%%%%%%%%%%%%%%%%%%%%%%%%%%

%%%%%%%%%%%%%%%%%%%%%%%%%%%%%%%%%%%%%%%%%%%%%%%%%%%%%%%%%%%%%
%%%%%%%%%%%%%%%%%%%%%%%%%%%%%%%%%%%%%%%%%%%%%%%%%%%%%%%%%%%%%
%%%%%%%%%%%%%%%%%%%%%%%%%%%%%%%%%%%%%%%%%%%%%%%%%%%%%%%%%%%%%
%%%%%%%%%%%%%%%%%%%%%%%%%%%%%%%%%%%%%%%%%%%%%%%%%%%%%%%%%%%%%
%%%%%%%%%%%%%%%%%%%%%%%%%%%%%%%%%%%%%%%%%%%%%%%%%%%%%%%%%%%%%
%%%%%%%%%%%%%%%%%%%%%%%%%%%%%%%%%%%%%%%%%%%%%%%%%%%%%%%%%%%%%
\section{\label{sec:conclusion}
        Conclusion}

The latest observational results from the ACT collaboration suggest a higher scalar spectral index, reported as $n_s = 0.9743 \pm 0.0034$, which is notably larger than the earlier estimate from the Planck mission~\cite{Planck:2018jri}. This upward shift in $n_s$ poses a significant challenge to many inflationary models that were previously consistent with Planck data. As a result, several inflationary models, are now disfavored under the new ACT observations.

In this work, we investigated inflationary dynamics within the framework of EGB gravity, incorporating a fractional scalar potential. The EGB theory introduces a quadratic curvature correction to the standard Einstein–Hilbert action through a non-minimal coupling between the scalar field and the Gauss–Bonnet term. We analyzed two distinct forms of the coupling function, namely $\xi(\phi) \propto \tanh(\xi_2 \phi)$ and $\xi(\phi) \propto \exp(\xi_2 \phi)$. Owing to the modified field equations arising from the Gauss–Bonnet interaction, the model offers a viable framework for accommodating inflationary potentials for the different values of $n$ that are otherwise ruled out by the latest observational constraints from ACT.
 
 We employed the slow-roll approximation to evaluate the model’s predictions for the scalar spectral index $n_s$ and the tensor-to-scalar ratio $r$. The resulting $r$–$n_s$ values were compared with observational bounds by plotting them on the $r$–$n_s$ plane for the different fractional values of $n$ of the scalar potential, constrained by ACT data.
The calculated values of $n_s$ and $r$~(for numerical values, see Table.~\ref{tab:coupling-clear}) fall within the $1\sigma$ confidence region of the ACT observations, indicating good agreement between the theoretical model and data.  

To further analyze the model, we explore the parameter space of the coupling coefficients $\xi_1$ and $\xi_2$. For the case of $\tanh$-type coupling, it was observed that higher values of $\xi_1$ and $\xi_2$  results in the better agreement with the bounds imposed on the $r-n_s$ by the ACT observations (see Fig.~\ref{tanxi}). A similar trend can be seen for the other coupling function ($\rm Exp$). Where higher values of $\xi_1$ and $\xi_2$ results in observationally viable values of $r-n_s$ (see Fig.~\ref{expxi}). Finally, we examined the running of the scalar spectral index ($\alpha_s$) for both type of coupling function mentioned in Eq.~\eqref{coupling_function} and the results remains consistent with the observational bounds.

At the final stage of our analysis, we examined the reheating phase by plotting the variation of the reheating e-fold number $N_{\text{re}}$ and the reheating temperature $T_{\text{re}}$ with respect to the scalar spectral index $n_s$. The results indicated that, in order to simultaneously satisfy the ACT observational constraint $n_s = 0.974$ and the bounds on the reheating temperature, the equation-of-state parameter during reheating must lie in the range $\omega_{\text{re}} > 1/3$. Furthermore, it was observed that increasing $\omega_{\text{re}}$ leads to a decrease in the number of reheating e-folds and an increase in the reheating temperature.

\vspace{0.3cm}
\noindent
In our future work, we aim to investigate the formation of Primordial Black Holes (PBHs) and the generation of Scalar-Induced Gravitational Waves (SIGWs) within the context of modified gravity theories~\cite{Kawai:2021edk,Yogesh:2025hll,Gangopadhyay:2023qjr,Gangopadhyay:2021kmf}.

\iffalse
It should also be noted that the model has also been considered using the new slow-roll approximations, proposed in~\cite{Pozdeeva:2024ihc}. The result shows no noticeable difference from the standard slow-roll approximations of the EGB gravity. 
\fi

\section*{Acknowledgements}
The authors thank Yogesh for discussing the idea for this project and his continued suggestions and insightful discussions throughout its development.

%The authors would like to thank J. Kim for the useful discussions and suggestions. This work is supported by the National Natural Science Foundation of China under Grants No. 12275238, the Zhejiang Provincial Natural Science Foundation of China under Grants No. LR21A050001 and No. LY20A050002, and the National Key Research and Development Program of China under Grant No. 2020YFC2201503.

%It was determined that for the valid range of the coupling free parameter $0.0012 < \xi_2 \leq 0.01$ the $r-n_s$ values stand in the $1\sigma$ region of the P-ACT-LB-BK18, showing that the Fractional potential in the EGB gravity theory can adequately satisfy the constraint provided by the ACT data. 

%\nocite{*}
\bibliographystyle{apsrev4-1}
\bibliography{EGB_ACT}% Produces the bibliography via BibTeX.

\end{document}